\documentstyle[11pt,newpasp,twoside]{article}
\def\gsim{\;\rlap{\lower 2.5pt
\hbox{$\sim$}}\raise 1.5pt\hbox{$>$}\;}
\def\lsim{\;\rlap{\lower 2.5pt
   \hbox{$\sim$}}\raise 1.5pt\hbox{$<$}\;}   

\def\edcomment#1{\iffalse\marginpar{\raggedright\sl#1\/}\else\relax\fi}
\reversemarginpar

\begin{document}
\title{THE QUIESCENCE OF DWARF NOVAE AND X-RAY TRANSIENTS}
 \author{Kristen Menou}
\affil{Chandra Fellow, Princeton University, Department of Astrophysical Sciences, Princeton, NJ 08540, USA}

\begin{abstract}
Our understanding of the structure and properties of accretion disks
in quiescent Dwarf Novae and X-ray Transients is very
limited. Observations during quiescence challenge some of the standard
Disk Instability Model (DIM) predictions. Our ignorance of the nature
of viscosity may be the main source of uncertainties in quiescent disk
models. A ``layered accretion'' alternative to the DIM, in which
accretion proceeds via X-ray ionized surface layers, illustrates the
magnitude of these uncertainties. A detection of molecular hydrogen in
quiescent disks may provide direct constraints on the nature of
viscosity.

\end{abstract}

\section{Introduction}

Dwarf Novae (DN) and Soft X-ray Transients (SXTs) spend most of their
lifetime in quiescence. According to the Disk Instability Model (DIM),
mass builds up in an unsteady, neutral disk during this phase, until
an outburst is triggered (Hameury, this volume; Lasota 2001).

While observations support our theoretical understanding of disk
accretion during outburst, the quiescence phase is only poorly
understood. In outburst, the presence of a hot, quasi-steady disk
expected in the DIM picture is supported by eclipse mapping studies in
DN (see, e.g., Horne 1993 for a review) and X-ray spectral fits in
SXTs (e.g. Tanaka \& Shibazaki 1996). There is every reason to believe
that angular momentum transport in these disks is due to MHD
turbulence resulting from the Magneto-Rotational Instability (MRI;
Balbus \& Hawley 1991, 1998; Balbus, this volume).

Observations in quiescence, however, are mostly puzzling, if not
challenging the predictions of the DIM (see \S2). The
nature of viscosity in the neutral, quiescent disks is also uncertain,
if not unknown (see \S4). In these conditions, it is worth
questioning the relevance of the DIM for describing the quiescence of
DN and SXTs.
 
A brief summary of observational data on quiescent disks is first
given in \S2. A critical discussion of the reliability of various
features in the DIM then follows in \S3. The main factor limiting our
understanding of quiescence in DN and SXTs may be our ignorance of the
nature of viscosity in quiescent disks, as is discussed in \S4. To
illustrate this point, an alternative to the standard DIM picture is
proposed in \S5 in the form of a layered accretion toy model. The
advantages and possible difficulties of this model are discussed in
\S6 before concluding in \S7.

\section{Observations During Quiescence}
\subsection{High Energy}
\subsubsection{Soft X-ray Transients (SXTs) --} 

Lasota (1996) pointed out that if quiescent disks in SXTs were
accurately described by the DIM, accretion onto the compact object
should proceed at a rate $\ll 10^{10}$~g~s$^{-1}$, i.e. many orders of
magnitude smaller than inferred from X-ray observations. In addition,
at such a low rate, a thin accretion disk is not expected to emit a
substantial amount of X-rays. This has led to the general belief that
there cannot be a disk extending all the way down to the compact
object in quiescent SXTs as would be expected in the DIM (see, e.g.,
Esin, McClintock \& Narayan 1997).  Wheeler (1996) also pointed out
that the optical light from quiescent black hole SXTs (excluding the
contribution from the companion star) corresponds to black body
temperatures in excess of $10^4$~K (see, e.g., Esin et al. 1997),
which is too hot for the neutral disk expected in the DIM. It appears
possible, however, that the quiescent optical luminosity of SXTs is
powered by bright spot emission (Menou \& McClintock 2001).

Nonetheless, the notion that a disk (evidenced by broad, double-peaked
emission lines; e.g. Orosz et al. 1994) accumulates mass during
quiescence to later power luminous outbursts seems quite reasonable.

\subsubsection{Dwarf Novae (DN) --}

Brown, Bildsten \& Rutledge (1998) and Bildsten \& Rutledge (2000; but
see Lasota 2000) have proposed alternatives to accretion for powering
the quiescent X-ray luminosity of SXTs. In quiescent DN, however,
the shapes of X-ray eclipses in several quiescent DN shows that hard
X-ray emission originates from the vicinity of the white dwarf (WD;
Wheatley, this volume), making any alternative to boundary layer
emission (and therefore accretion) rather implausible.

A significant increase of the
accretion rate onto the compact object over the quiescence
period is predicted by the DIM. 
This prediction is challenged by observations which show
instead a (slightly) decreasing hard X-ray luminosity (van der Woerd
\& Heise 1987; Verbunt, Wheatley \& Mattei 1999).

The decreasing UV fluxes observed in several quiescent DN could also
be interpreted as challenging the DIM, but given that in some cases
the UV light is clearly dominated by WD or accretion belt cooling
(see, e.g., Sion 1999 for a review), this interpretation is subject
to caution.

\subsection{Optical}

One also expects the optical luminosity of the disk to increase during
quiescence in the DIM. van Amerongen, Kuulkers \&
van Paradijs (1990), for instance, isolated the disk contribution in
the quiescent lightcurve of Z Cha.  They found that the disk
contribution is nearly constant over the quiescence period, in
contradiction with DIM predictions.

Eclipse mapping studies have revealed that quiescent disks in DN have
brightness temperature profiles rising inward from $\sim 5000$~K to
$\sim 8000$~K, with slopes significantly shallower than $R^{-3/4}$
(the expected scaling for the effective temperature of a steady-state
disk, as observed in outburst; see, e.g., Horne 1993).

The colors and strong emission lines of quiescent disks in DN have
long suggested emission from optically-thin gas (see, e.g., Robinson,
Marsh \& Smak 1993 for a review). Modeling this emission as an
isothermal slab has systematically resulted in inferred values of the
viscosity parameter $\alpha$ of several hundreds, much in excess of a
reasonable $\alpha < 1$. It is therefore unclear whether these models,
which require unsteady disks to reproduce the observed brightness temperature
profiles, really support the DIM.

In addition, this type of models marginally reproduces the strength of
the observed H emission lines, but is unable to account for the strong
He lines observed (Robinson et al. 1993). Including high-energy
irradiation in radiative models of quiescent DN disks seems required
to reproduce these He lines (e.g. Patterson \& Raymond 1985; Ko et
al. 1996) and could result in large structural changes (e.g. vertical
temperature inversion) with consequences for the emission that remain
largely unexplored to date (Robinson et al. 1993).

\section{Consequences for the DIM}

The entirely optically-thin quiescent disk described by the isothermal
slab model appears inconsistent with the DIM. Indeed, while the DIM
would predict a typical surface density $\Sigma \sim 100$~g~cm$^{-2}$
at a radius of $\sim 10^{10}$~cm in a quiescent disk (e.g. Hameury et
al. 1998), the isothermal slab model requires much smaller values of
$\Sigma$ for the slab to be optically-thin, typically
$<1$~g~cm$^{-2}$.

The best attempt to date to compare observations of quiescent disks
with expectations from the DIM is that of Idan et al. (1999). These
authors calculated the emission from an unsteady quiescent disk (with
an accretion rate $\dot M$ roughly $\propto R^3$, as predicted in the
DIM) with Shaviv--Wehrse models (which include a treatment of the
optically thin regions of the disk).  By calibrating their models to
the observed colors and inferred accretion rate of the DN HT Cas
during quiescence, they concluded that a viscosity parameter $\alpha >
1$ is required to fit the observations. Idan et al. emphasize that
this value is inconsistent with the typical value $\alpha_{\rm cold}
\sim 0.01$ inferred from modeling DN recurrence times with the DIM.

To date, the most realistic radiative models of quiescent disks (as
described by the DIM) are therefore not internally consistent. It is
worth noting, however, that the models of Idan et al. also neglect the
role of high-energy irradiation (required to reproduce the observed He
lines). Although this may not solve the $\alpha > 1$ problem, there is
in principle still room for improvement.  Nonetheless, the
discrepancies between the quiescent variations of the hard X-ray and
optical luminosities expected in the DIM and the observations
seriously challenge the model.

Because of this, it is worth considering which of the DIM features for
quiescent disks are reliable. The idea that mass accumulates during
quiescence to power the luminous disk outbursts is certainly supported
by observations in a general sense. Another feature of the DIM that
appears reliable is the initial quiescent profile of surface density
$\Sigma$ in the disk (immediately following the propagation of a
cooling front), because it is independent of the adopted value of
$\alpha_{\rm cold}$ when $\alpha_{\rm cold} \ll \alpha_{\rm hot}$
(Menou, Hameury \& Stehle 1999). The subsequent disk evolution and
most of the other quiescent disk features in the DIM are questionable,
however, because of our ignorance of the nature of viscosity in these
disks. In particular, the global stability criterion on the mass
accretion rate, $\dot M_{\rm crit}^- (R)$, used to determine when an
outburst is triggered may not be reliable.

\section{Nature of Viscosity}

\subsection{MHD Turbulence}

The MRI most likely operates in hot disks during outbursts, leading to
efficient MHD-turbulent transport. The situation for quiescent disks
is much less clear, however. Non-ideal MHD effects become important in
a weakly-ionized gas. Resistive diffusion should be the dominant such
process in quiescent disks (Gammie \& Menou 1998). Hall effects
contribute to a smaller extent, given the typical densities $n \sim
10^{18}$~cm$^{-3}$ expected (and making the reasonable assumption that
gas pressure dominates over magnetic pressure; Balbus \& Terquem
2001).

To date, various estimates and numerical simulations have suggested
that MHD turbulence dies away in quiescent disks (Gammie \& Menou
1998; Fleming, Stone \& Hawley 2000; Menou 2000). These works have
neglected Hall effects, however. Future work including Hall effects
will verify if MHD turbulence indeed shuts off in quiescence or if a
residual, low-level of MHD turbulence is expected (Balbus, this
volume).

\subsection{The Role of Molecular Hydrogen}
It is worth emphasizing here that the suppression of MHD turbulence in
quiescent disks seems to require hydrogen to be in molecular phase,
simply because dissociative recombination is several orders of
magnitude more efficient than atomic hydrogen recombination (see
\S5.1). In that sense, a direct detection of molecular hydrogen would
likely provide strong constraints on the nature of viscosity in
quiescent disks. Band diagnostics may reveal the gas temperature and
density in the disk, which would allow a direct, {\it in situ} measure
of the strength of non-ideal MHD effects.

\subsection{Other ``Viscosity'' Mechanisms?}

Numerical simulations of Keplerian disks, which show the development 
of turbulence and efficient transport in the magnetized 
case but not in the unmagnetized case, support the idea that disks are
hydrodynamically stable (Hawley, Balbus \& Winters 1999; see Richard
\& Zahn 1999 for a different view). In addition, while spiral density
waves may play a role in determining outburst recurrence times (Menou
2000), they are not expected to be present in quiescent disks because
of the inefficient coupling of the companion tidal field with a cold
disk (Boffin, this volume). Finally, disk self-gravitation is probably
irrelevant for quiescent disks in DN and SXTs given the relatively
small disk masses involved and the disk floor temperature expected,
e.g., from irradiation by the companion star.  In other words, if MHD
turbulence dies away in (the bulk of) a quiescent disk, there may be
no other mechanism for transporting angular momentum. This is the main
motivation for studying layered accretion. (Of course, layered
accretion could still be relevant even if there is a residual
viscosity in the bulk of quiescent disks).

\section{Layered Accretion Toy Model}

The layered accretion model presented here is inspired by Gammie's
(1996) proposal for T-Tauri disks. In Gammie's picture, cosmic-rays
penetrate a layer $\sim 100$~g~cm$^{-2}$ deep and provide non-thermal
ionization allowing MHD turbulence to develop and accretion to proceed
in disk surface layers. In between the two active layers lies a dead
zone, where no accretion occurs if no viscosity mechanism operates.

One easily shows that, in DN (and SXT) quiescent disks, cosmic ray
ionization is rather inefficient because of the much larger densities
involved. On the other hand, hard X-ray emission, typically at a
level of $10^{31}$~erg~s$^{-1}$ (and Bremsstrahlung temperatures $kT
\sim 1-10$~keV) may provide the required ionization level for MHD
turbulence to develop in surface layers. (Irradiation also constitutes
an extra source of heating for the disk, but it is generally
negligible when compared to viscous dissipation in the active layers).

The cross-section of a $5-7$~keV X-ray photon in a solar-composition
material approximately corresponds to $\sim 1$~g~cm$^{-2}$, so that
this value is chosen for the normalization of the active layer surface
density, $\Sigma_a$ (ignoring geometrical effects). Softer X-ray
photons have a larger cross-section and therefore penetrate less deep
inside the disk (and vice-versa; Morrison \& McCammon 1983). Note that
a layer of $\sim 1$~g~cm$^{-2}$ with a temperature of a few $1000$~K
(see solution below) should be optically-thin.

\subsection{Equations}

Mass and angular momentum conservation in the steady-state, active
layers yields
\begin{equation}
\dot M=6 \pi R^{1/2} \frac{\partial }{\partial R} \left( 2 \Sigma_a
\nu R^{1/2}\right),
\end{equation}
where $\dot M$ is the summed accretion rate in the 2 active layers,
$\nu$ is the kinematic viscosity in these layers and $R$ is the
distance from the central object.  Conservation of the
viscously-dissipated energy yields
\begin{equation}
\frac{9}{4} \nu \Sigma_a \Omega_k^2=\sigma T_{\rm eff}^4,
\end{equation}
where $\Omega_k$ is the Keplerian angular rotation speed and $T_{\rm
eff}$ is the effective temperature of the active layers.  The
radiative transfer (assumed optically-thin) is accounted for in the
simplest way:
\begin{equation}
\tau T_c^4=T_{\rm eff}^4,~\tau=\Sigma_a \kappa_p,
\end{equation}
where $T_c$ is the temperature of the active layers, $\tau$ is the
optical thickness of each active layer and $\kappa_p$ a Planck-mean
opacity.  Finally, a standard Shakura--Sunyaev $\alpha$--viscosity is
assumed:
\begin{equation}
\nu=\alpha c_s^2/\Omega_k.
\end{equation}
The above equations are closely related to the standard steady-state
thin disk equations with the additional assumption of optically-thin
radiative transfer.

The flux irradiating the disk is taken as
\begin{equation}
F_x \approx \frac{L_x}{4 \pi R^2}\frac{H}{R}, 
\end{equation}
where $H$ is the disk geometrical thickness. The ratio $H/R$ should
reasonably approximate the irradiation geometry if the hard X-rays
originates in a hot, optically-thin boundary layer of size comparable
to the WD. The volumic ionization rate is therefore
\begin{equation}
\xi_i \approx \frac{L_x}{4 \pi R^3 E_i}\frac{H}{H_a},
\end{equation} 
where $H_a$, the active layer thickness, has been introduced for
generality and $E_i \approx 37$~eV is the typical energy required for
secondary electron generation (Glassgold, Najita \& Igea 1997). The
volumic recombination rate for a predominantly-atomic hydrogen gas is
$\xi_r \approx 6.68 \times 10^{-13} n_e^2 (T_{c}/3000~{\rm
K})^{-0.8}$~cm$^3$~s$^{-1}$ ($n_e$ is the electron number density); it
becomes much faster for a predominantly-molecular hydrogen gas, with
$\xi_r \approx 8.7 \times 10^{-6} n_e^2 T_c^{-0.5}$~cm$^3$~s$^{-1}$
(dissociative recombination).

The magnetic Reynolds number in the active layers can be calculated
according to
\begin{equation}
Re_M \equiv \frac{c_s H_a}{\eta}, \eta=\frac{c^2 m_e \nu_{en}}{4 \pi
n_e e^2},
\end{equation}
where $c_s$ is the sound speed in the active layers, $\eta$ is the
resistivity, proportional to the electron-neutral collision frequency,
$\nu_{en}$, and other symbols have their usual meaning (Gammie \&
Menou 1998). The equivalent dimensionless number measuring the
strength of ambipolar diffusion is
\begin{equation}
Re_A= \frac{\nu_{ni}}{\Omega_k},
\end{equation}
proportional to the ion-neutral collision frequency, $\nu_{ni}$ (Menou
\& Quataert 2001).

\subsection{Solution}
The above structural equations are solved assuming, instead of a
constant $\dot M$ with radius like in the Shakura-Sunyaev solutions, a
constant surface density with radius, $\Sigma_a$ (in g~cm$^{-2}$), in
the active layers.

\subsubsection{Structure--}
The temperature, effective temperature and accretion rate of the
active layers scale as
\begin{equation}
T_c=3357~{\rm K}~ \left( \frac{\kappa_p}{0.1} \right)^{-1/3}
\alpha_{0.1}^{1/3} m_1^{1/6} R_{10}^{-1/2},
\end{equation}
\begin{equation}
T_{\rm eff}=1888~{\rm K}~ \Sigma_a^{1/4} \left( \frac{\kappa_p}{0.1}
\right)^{-1/12} \alpha_{0.1}^{1/3} m_1^{1/6} R_{10}^{-1/2},
\end{equation}
\begin{equation}
\dot M=1.4 \times 10^{14}~{\rm g~s^{-1}}~ \Sigma_a \left(
\frac{\kappa_p}{0.1} \right)^{-1/3} \alpha_{0.1}^{4/3} m_1^{-1/3}
R_{10},
\end{equation}
where an arbitrary $\kappa_p=0.1$ and $\alpha_{0.1}=\alpha / 0.1=1$
have been assumed, $m_1$ is the central object mass in solar units and
$R_{10}$ is the distance from the accretor in units of $10^{10}$~cm.
A typical mass density in the active layers is $\rho=\Sigma_a/H_a \sim
10^{-8}$~g~cm$^{-3}$. Assuming that no viscous dissipation occurs in
the dead zone, it should be isothermal with $T \sim T_c$, if
conduction between active and dead layers is efficient (i.e. at
thermal equilibrium). In the limit where conduction is inefficient,
the dead zone (assumed optically-thick) should have $T \sim T_{\rm
eff}$ since it absorbs and reradiates about half the flux emitted by
the optically-thin active layers.

To within an order of magnitude or so, the above accretion rate, at $R
\simeq 10^9$~cm (an appropriate disk inner radius in the case of
accretion onto a WD), is sufficient to power $10^{31}$~erg~s$^{-1}$ in
a quiescent DN. Note that $\dot M$ does not depend on $L_x$, which
guarantees stability against fluctuations of $L_x$ as often observed
in quiescent DN. ($L_x$ determines the quality of gas-field coupling
in the active layers, however, via Eq.~[6]). The scaling of $\dot M$
with $\Sigma_a$ implies instead a sensitivity to the spectrum of
irradiating photons.

\subsubsection{Ionization Properties --} 
The values of the electron density, ionization fraction, magnetic
Reynolds number and equivalent for ambipolar diffusion, in the active
layers, scale as
\begin{equation}
n_e=1.4 \times 10^{11}~{\rm cm^{-3}}~ \left( \frac{T_c}{3000~{\rm
K}}\right)^{0.4} \left( \frac{L_x}{10^{31}~{\rm
erg~s^{-1}}}\right)^{1/2} \left( \frac{H_a}{H}\right)^{-1/2}
R_{10}^{-3/2},
\end{equation}
\begin{equation}
x_i=10^{-5}~ \Sigma_a^{-1} m_1^{-1/2} \left( \frac{T_c}{3000~{\rm
K}}\right)^{0.9} \left( \frac{L_x}{10^{31}~{\rm
erg~s^{-1}}}\right)^{1/2} \left( \frac{H_a}{H}\right)^{1/2},
\end{equation}
\begin{equation}
Re_M=1.5 \times 10^4~ \Sigma_a^{-1} m_1^{-1} \left(
\frac{T_c}{3000~{\rm K}}\right)^{1.4} \left( \frac{L_x}{10^{31}~{\rm
erg~s^{-1}}}\right)^{1/2} \left( \frac{H_a}{H}\right)^{1/2}
R_{10}^{3/2},
\end{equation}
\begin{equation}
Re_A= 700~ m_1^{-1/2} \left( \frac{T_c}{3000~{\rm K}}\right)^{0.4}
\left( \frac{L_x}{10^{31}~{\rm erg~s^{-1}}}\right)^{1/2} \left(
\frac{H_a}{H}\right)^{-1/2}, 
\end{equation}
where a fixed temperature value, $T_c=3000$~K, has been used in these
estimates for simplicity. A comparison of $Re_M$ and $Re_A$ above to
existing estimates of the critical values ($Re_{M,c} \sim 10^4$,
$Re_{A,c} \sim 100$; Fleming et al. 2000; Hawley \& Stone 1998) below
which MHD turbulence cannot be self-sustained suggests only marginal
gas-field coupling in the active layers.

Note that at $T_c \sim 3000$~K, the dead zone ($\rho \sim
10^{-6}$~g~cm$^{-3}$) is predominantly molecular, while the active
layers ($\rho \sim 10^{-8}$~g~cm$^{-3}$) are predominantly atomic
(see, e.g., Lenzuni, Chernoff \& Salpeter 1991). This justifies the
use of the slower recombination rate $\xi_r$ for an atomic gas in the
solution above. If hydrogen were predominantly molecular in the active
layers, the value of $Re_{M}$ would be several orders of magnitude
smaller than indicated above.

\subsection{Mass Accumulation}

\subsubsection{Growth Timescale --} 
The accretion rate in the active layers, $\dot M$, decreases with
radius, so that mass accumulates in the dead zone as a function of
time, according to
\begin{equation}
\frac{\partial \Sigma_d}{\partial t} \equiv \frac{1}{2 \pi R}
\frac{\partial \dot M}{\partial R}.
\end{equation}
Approximating the initial quiescent density profile in the dead zone
as $\Sigma_d \sim \Sigma_{\rm min}~(\alpha=0.1) \propto R_{10}^{1.11}$
(e.g. Hameury et al. 1998), the timescale over which mass build ups in
the dead zone is then
\begin{equation}
\frac{\Sigma_d}{\partial \Sigma_d / \partial t}= 7.2~{\rm yrs}~
\Sigma_a^{-1} m_1^{-0.04} \alpha_{0.1}^{-2.1} \left(
\frac{\kappa_p}{0.1} \right)^{1/3} R_{10}^{2.11},
\end{equation}
which corresponds to the interesting values of $\sim 20$ days at $\sim
10^9$~cm (disk inner radius in the DN case) and $\sim 30$ years at $
\sim 2 \times 10^{10}$~cm (a typical disk transition radius in the
quiescent BH SXT models of Esin et al. 1997). These timescales are
comparable to the typical recurrence times of DN and BH SXTs,
respectively. Mass could also accumulate in the disk outer regions for
which Eq.~(11) indicates accretion rates smaller than typical mass
transfer rates (at least in some systems).  Note also that
$\Sigma_{\rm min}$ becomes $< 1$~g~cm$^{-2}$ (the typical penetration
scale of photoionizing X-ray photons) at radii smaller than $\sim 3
\times 10^8$~cm. Consequently, a fully active, steady-state disk
(i.e. without a dead zone or mass accumulation) is expected at small radii
(at least in SXTs).

\subsubsection{How to Trigger Outbursts? --}
Mass accumulation by itself does not result in outbursts. If there is
a residual viscosity in the ``dead'' zone, however, the viscous torque
and viscous dissipation, both $\propto \Sigma_d$, will build up with
time ($\alpha_{\rm dead}$ could be $<<1$). In these conditions, the
central temperature in the likely optically-thick, ``dead'' zone will
also increase with time, potentially resulting in sufficient thermal
ionization at late times to initiate MHD turbulence and a global disk
outburst. An interesting candidate for the residual viscosity process
is a Reynolds (i.e. hydrodynamical) stress induced in the ``dead''
zone by the MHD turbulence in the active layers, as found by Fleming
\& Stone (2001) in their numerical simulations of layered accretion
for T-Tauri disks.

\section{Discussion}

As a model of quiescent accretion in DN (and possibly SXTs), the
layered accretion toy model outlined above has several advantages and
shortcomings. With $\Sigma_a \sim 1$~g~cm$^{-2}$ and $T_c \sim$ a few
$10^3$~K, optically-thin gas emission is expected from the active
layers. The profile of effective temperature ($T_{\rm eff}(R) \propto
R^{-1/2}$) is flatter than for a steady-state disk ($R^{-3/4}$) and
probably in reasonable agreement with inferences from quiescent eclipse
maps. Angular momentum transport is solely provided by the well
understood MRI (MHD turbulence) in this scenario. Contrary to the DIM,
no increase in $\dot M$ during quiescence is expected in this scenario
(though no decrease is expected either). The important high-energy
irradiation of quiescent disks is a crucial component of the model
which is often ignored in other descriptions of quiescent
disks. Finally, various scalings in the layered accretion solution
are, at an order of magnitude level, in agreement with observationally
inferred values, such as the quiescent accretion rate onto WDs in DN.
One certainly cannot ask much more from such an idealized, toy model.

The extremely crude optically-thin radiative transfer treatment or the
1--zone approximation are examples of obvious oversimplifications in
the model. The only marginal MHD coupling of the active layers can
also be considered as a weak point of the model (though Hall effects
may help gas-field coupling in this low-density case). Fine-tuning may
be required to guarantee, e.g., the predominantly-atomic nature of the
active layers on top of a predominantly-molecular dead zone. It is
also unclear how well the model applies to SXTs since, for instance,
disk self-irradiation could be much less in BH SXTs than irradiation
by the central WD in DN (though reprocessing by a hot extended medium
such as an ADAF may help). Finally, because the simulations of Fleming
\& Stone show that the magnitude of induced-viscosity in the dead zone
decreases for larger ratios of dead-zone to active-layer mass, an
additional, yet unidentified viscosity mechanisms may be required, in
the end, to trigger outbursts in the dead zone.

Nonetheless, the layered accretion toy model has the important
advantage of offering a plausible alternative to the quiescent disk
structure predicted by the DIM and challenged by observations of
quiescent DN. At the very least, the layered accretion scenario should
be used as a measure of our poor understanding of quiescent disks due
to our ignorance of the nature of viscosity in these disks. In that
respect, it is quite significant that the typical value of
$\alpha_{\rm cold} \sim 0.01$ inferred from DN recurrence times does
not necessarily measure an actual efficiency of angular momentum
transport in the layered accretion scenario but rather corresponds to
the timescale for mass accumulation in the dead zone ($\alpha$ being
presumably $\sim 0.1$ in well-coupled active layers and possibly $<<
0.01$ in the dead zone).

\section{Conclusion}

The unknown nature of viscosity in the quiescent disks of DN and SXTs
results in considerable uncertainty on the structure and properties of
these disks. Layered accretion appears as a plausible alternative to
the standard DIM picture, although additional work is clearly required
to validate this scenario. 

\vspace{0.3cm}

{\it The author thanks J. Stone for communicating results prior 
to publication, J.-P. Lasota for comments on the manuscript 
and the Center for Astrophysical Sciences at
Johns Hopkins University for hospitality. 
Support for this work was provided by NASA through
Chandra Fellowship grant PF9-10006 awarded by the Smithsonian
Astrophysical Observatory for NASA under contract NAS8-39073.}

\end{document}